\documentclass[12pt]{article}
\usepackage{epsfig}

\def\beq{\begin{equation}}
\def\eeq#1{\label{#1}\end{equation}}
\def\eeqn{\end{equation}}

\def\beqa{\begin{eqnarray}}
\def\eeqa#1{\label{#1}\end{eqnarray}}
\def\eeqan{\end{eqnarray}}

\let\bar=\overbar

\def\bra#1{\left\langle{ #1} \right|}
\def\ket#1{\left| {#1} \right\rangle}

\def\Dslash{\not{\hbox{\kern-4pt $D$}}}
\def\dslash{\not{\hbox{\kern-2pt $\del$}}}

\def\msb{{\bar{\ssstyle M \kern -1pt S}}}

\def\beq{\begin{equation}}
\def\eeq{\end{equation}}
\def\bea{\begin{eqnarray}}
\def\eea{\end{eqnarray}}
\def\nn{\nonumber}
\def\roughly#1{\mathrel{\raise.3ex\hbox
{$#1$\kern-.75em\lower1ex\hbox{$\sim$}}}}

\def\bra#1{\left\langle #1\right|}
\def\ket#1{\left| #1\right\rangle}
\def\bd{B_d^0}

\def\bs{B_s^0}
\def\bsbar{{\bar B}^0_s}
\def\sss{\scriptscriptstyle}
\def\ks{K_{\sss S}}
\def\kbar{{\bar K}^0}
\def\kstarbar{{\bar K}^*}
\def\Puc{{\cal P}_{uc}}
\def\Ptc{{\cal P}_{tc}}

\def\ZI{Z_{\sss I}}
\def\ZR{Z_{\sss R}}

\def\plb#1#2#3{{\it Phys.\ Lett.} {\bf #1B}, #3 (#2)}
\def\prd#1#2#3{{\it Phys.\ Rev.} {\bf D#1}, #3 (#2)}
\def\newprd#1#2#3{{\it Phys.\ Rev.} {\bf D#1}: #3 (#2)}

\def\prl#1#2#3{{\it Phys.\ Rev.\ Lett.} {\bf #1}, #3 (#2)}

\def\Title#1{\begin{center} {\Large {\bf #1} } \end{center}}

\begin{document}

\begin{flushright}
UdeM-GPP-TH-02-97\\
\end{flushright}

\Title{Measuring $\alpha$ using $B \to K^{(*)} {\bar K}^{(*)}$
Decays\footnote{Talk given by David London at {\it Flavor Physics and
CP Violation (FPCP)}, Philadelphia, PA, USA, May 2002}}

\bigskip\bigskip


\begin{raggedright}  

{\it Alakabha Datta and David London\index{xxx} \\
Laboratoire Ren\'e J.-A. L\'evesque \\
Universit\'e de Montr\'eal \\
C.P. 6128, succ.\ centre-ville \\
Montr\'eal, QC, CANADA H3C 3J7}
\bigskip\bigskip
\end{raggedright}

The {\it raison d'\^etre} for measuring CP violation in the $B$ system
is to test the standard model (SM) \cite{CPreview}. As always, the
hope is that we will find evidence for the presence of new physics.

There are many signals of new physics, but only a handful are of the
``smoking-gun'' variety, i.e.\ they are free of hadronic
uncertainties, and hence independent of theoretical input. All of
these rely on the measurement of CP-violating rate asymmetries in $B$
decays. They include
\begin{itemize}

\item $\bd(t)\to \Psi \ks$ vs.\ $\bd(t) \to \phi \ks$. Both of these
decay modes probe the CP phase $\beta$ within the SM.

\item $B^\pm \to D K^\pm$ vs.\ $\bs(t) \to D_s^\pm K^\mp$. Similarly,
both of these modes can be used to measure $\gamma$.

\item $\bs(t) \to \Psi \phi$. The CP asymmetry for this decay is
expected to vanish within the SM (to a good approximation).

\end{itemize}
In all cases, any deviation from the SM predictions indicates the
presence of new physics.

However, these signals have something in common: they are all
sensitive to new physics in the $b\to s$ flavour-changing neutral
current (FCNC). In the first case, the new physics enters in the $b\to
s$ penguin amplitude, while in the last two cases, it affects
$\bs$--$\bsbar$ mixing.

This then begs the question: are there clean probes of new physics in
the $b\to d$ FCNC? However, the answer to this is {\it no}.

To see this, consider the $b\to d$ penguin amplitude, which can be
written as
\beq
A = P_u \, V_{ub}^* V_{ud} + P_c \, V_{cb}^* V_{cd} + P_t \, V_{tb}^*
V_{td} ~,
\eeq
where the $P_i$ ($i=u,c,t$) represent the contributions from the
internal $i$-quark. Due to the unitarity of the
Cabibbo-Kobayashi-Maskawa (CKM) matrix, any one CKM combination can be
eliminated in terms of the other two. (Note that, unlike the $b\to s$
penguin amplitude, here all three combinations of CKM matrix elements
are of the same order.) There are thus three ways of writing this
amplitude:
\begin{enumerate}

\item $A = (P_u - P_c) V_{ub}^* V_{ud} + (P_t - P_c) V_{tb}^* V_{td}$,

\item $A = (P_c - P_u) V_{cb}^* V_{cd} + (P_t - P_u) V_{tb}^* V_{td}$,

\item $A = (P_u - P_t) V_{ub}^* V_{ud} + (P_c - P_t) V_{cb}^* V_{cd}$.

\end{enumerate}
In the first case, the relative weak phase between the two
contributions is $\alpha$, while in the second and third cases, it is
$\beta$ and $\gamma$, respectively. Therefore, there is an ambiguity
--- called the {\it CKM ambiguity} --- in the definition of the
relative weak phase in the $b\to d$ penguin amplitude. Because of
this, it is impossible to cleanly extract weak-phase information from
$b \to d$ penguins. Thus, in order to test for new physics in the
$b\to d$ FCNC, we need a theoretical assumption which will break the
CKM ambiguity \cite{LSS}. This fact will be important in what follows.

Consider the pure $b\to d$ penguin decay $\bd\to K^0 \kbar$. Its
amplitude can be written
\beq
A = P_u \, V_{ub}^* V_{ud} + P_c \, V_{cb}^* V_{cd} + P_t \, V_{tb}^*
V_{td} = \Puc \, e^{i\gamma} \, e^{i\delta_{uc}} + \Ptc \, e^{-i\beta}
\, e^{i\delta_{tc}} ~.
\eeq
Note that the magnitudes of the CKM combinations $V_{ub}^* V_{ud}$ and
$V_{tb}^* V_{td}$ have been absorbed into $\Puc$ and $\Ptc$. In this
parametrization, there are 4 theoretical parameters: $\Puc$, $\Ptc$,
$\Delta \equiv \delta_{uc} - \delta_{tc}$, and the CP phase $\alpha$.

Recall that the time-dependent decay rate for a $\bd$ to decay to a
final state $f$ can be written as
\beq
\Gamma(\bd(t) \to f) \sim X + Y \cos \Delta m t - \ZI \sin \Delta m t ~,
\eeq
where
\beq
X \equiv \frac{|A|^2 + |{\bar A}|^2}{2} ~,~~
Y \equiv \frac{|A|^2 - |{\bar A}|^2}{2} ~,~~
\ZI \equiv {\rm Im} \left( e^{-2i\beta} A^* {\bar A} \right) ~.
\eeq
What is important here is that there are only 3 experimental
observables. However, as noted above, there are 4 theoretical unknown
quantities describing the decay $\bd\to K^0 \kbar$. Thus, as expected,
one cannot extract any of these theoretical parameters. However, it is
always possible to express three of the unknowns in terms of the
fourth. In particular, a little algebra allows us to write
\beq
\Ptc^2 = \frac{\ZR \cos 2\alpha + \ZI \sin 2\alpha - X}{\cos 2\alpha -
1} ~,
\label{Ptcone}
\eeq
where $\ZR \equiv {\rm Re} \left( e^{-2i\beta} A^* {\bar A}
\right)$. (Note that $\ZR$ is not independent: $\ZR^2 = X^2 - Y^2 -
\ZI^2$.)

A similar analysis can be applied to the decay $\bd\to K^* \kstarbar$,
where $K^*$ represents any excited kaon, such as $K^*(892)$,
$K_1(1270)$, etc. In this case we can write
\beq
{\Ptc'}^2 = \frac{\ZR' \cos 2\alpha + \ZI' \sin 2\alpha - X'}{\cos
2\alpha - 1} ~,
\label{Ptctwo}
\eeq
where the primed observables correspond to the decay $\bd\to K^*
\kstarbar$. By combining Eqs.~(\ref{Ptcone}) and (\ref{Ptctwo}), we obtain
\beq
\frac{\Ptc^2}{{\Ptc'}^2} = \frac{\ZI \sin 2\alpha + \ZR
\cos 2\alpha - X} {\ZI' \sin 2\alpha + \ZR' \cos 2\alpha - X' } ~.
\label{Ptcratio}
\eeq
Note that the CKM information in $\Ptc$ and $\Ptc'$ (the magnitudes of
$V_{ub}^* V_{ud}$ and $V_{tb}^* V_{td}$) cancels in the ratio. The key
point here is the following: {\it if we knew the value of
$\Ptc^2/{\Ptc'}^2$, we could extract $\alpha$.}

We now turn to the analogous processes in the $\bs$ system. Consider
$\bs\to K^0\kbar$, which is a pure $b\to s$ penguin decay. Its
amplitude can be written
\beq
A^{(s)} = P_u^{(s)} \, V_{ub}^* V_{us} + P_c^{(s)} \, V_{cb}^* V_{cs}
+ P_t^{(s)} \, V_{tb}^* V_{ts} = \Puc^{(s)} \, e^{i\gamma} \,
e^{i\delta_{uc}^{(s)}} + \Ptc^{(s)} \, e^{i\delta_{tc}^{(s)}} ~.
\eeq
The difference between $b\to s$ penguins and $b\to d$ penguins is
that, here, $V_{ub}^* V_{us}$ is $O(\lambda^4)$ while $V_{tb}^*
V_{ts}$ is $O(\lambda^2)$. Thus, $\Puc^{(s)}$ is negligible compared
to $\Ptc^{(s)}$, so that the measurement of $B(\bs\to K^0\kbar)$ gives
$|\Ptc^{(s)}|$. Similarly, the measurement of $B(\bs\to K^*\kstarbar)$
gives $|\Ptc^{'(s)}|$.

We now make the claim that
\beq
{{\Ptc^{(s)}}^2 \over {\Ptc^{'(s)}}^2} = {\Ptc^2 \over {\Ptc'}^2} ~.
\label{claim}
\eeq
Note that the CKM matrix elements cancel in the ratios, so that this
is a relation among hadronic parameters. This theoretical assumption
breaks the CKM ambiguity. Thus, by combining this relation with that
in Eq.~(\ref{Ptcratio}), one can extract the CP phase $\alpha$ from
the pure penguin decays $B^0_{d,s} \to K^{(*)} {\bar K}^{(*)}$
\cite{BKKbar}. A similar method applies to non-CP-conjugate decays of
the form $K^0 {\bar K}^*$, $K^* \kbar$.

Of course, the precision with which $\alpha$ can be obtained depends
on the theoretical uncertainty in the above relation. As we will argue
below, the theoretical error is small, at most 5\% (and may well be
even smaller).

Before discussing the theoretical error, we briefly examine some of
the experimental considerations in putting this method to use. The
branching ratios for decays dominated by $b\to d$ penguins are
expected to be $O(10^{-6})$. Combined with the fact that $\bs$ decays
are involved, this suggests that this method is most appropriate for
hadron colliders. In addition, since the $K^{(*)}$ and ${\bar
K}^{(*)}$ mesons can be detected via their decays to charged $\pi$'s
and $K$'s, no $\pi^0$ detection is needed -- all that is necessary is
good $K/\pi$ separation. This is very important for most hadron
colliders.  (Of course, if $\pi^0$'s can actually be detected, this
will improve the prospects for using the method.) Finally, it should
be possible to trigger on a final-state $K^*$ at hadron colliders, so
that final states such as $K^* \kbar$, $K^0 {\bar K}^*$ and $K^* {\bar
K}^*$ will probably be favoured.

The method does have a potential weakness: $\alpha$ is extracted with
a 16-fold discrete ambiguity. However, this is not as serious as it
appears at first glance. First, the ambiguity can be reduced to 4-fold
by considering two different $K^{(*)} {\bar K}^{(*)}$ final
states. Examples of these include $K^0 \kbar$ and $K^* {\bar K}^*$,
$K^0 {\bar K}^*$ and $K^* \kbar$, or two different helicity states of
$K^* {\bar K}^*$. Second, we expect $P_u$ and $P_c$ in $b\to d$ penguins
to be at most 50\% of $P_t$. This implies that $\Puc/\Ptc < 0.5$ for
all decays.  By adding this theoretical constraint, the discrete
ambiguity can be reduced to 2-fold: $\alpha$, $\alpha+\pi$.

There are two other methods on the market for cleanly measuring
$\alpha$: (i) the isospin analysis of $B\to\pi\pi$ decays
\cite{isospin}, and (ii) the Dalitz-plot analysis of $B\to\rho\pi$
decays \cite{Dalitz}. Both of these methods have their problems. The
isospin analysis requires the measurement of $\bd\to\pi^0\pi^0$, whose
branching ratio may be quite small. And in the Dalitz-plot analysis,
one must understand the continuum background to $B\to\rho\pi$ decays
with considerable accuracy, and have a correct description of
$\rho\to\pi\pi$ decays.  Both of these issues may be difficult to
resolve. In light of this, the $B \to K^{(*)} {\bar K}^{(*)}$ method
could potentially give us the first reasonably clean measurement of
$\alpha$. However, regardless of which is first, a discrepancy in the
values of $\alpha$ obtained from 
these different methods
would point clearly to new physics
in the $b\to d$ penguin.

We now turn to a brief examination of the theoretical uncertainty in
Eq.~(\ref{claim}). We claim the equality of the double ratio of matrix
elements:
\beq
\frac{r_t}{r_t^*} \equiv \frac{\bra{K^0 \kbar} H_d \ket{\bd} /
\bra{K^0 \kbar} H_s \ket{\bs}}{\bra{K^* \kstarbar} H_d \ket{\bd} /
\bra{K^* \kstarbar} H_s \ket{\bs} } = 1 ~.
\eeq
The two decays in $r_t$ are related by U-spin (i.e.\ flavour $SU(3)$
symmetry), and similarly for $r_t^*$. We can therefore write
\beq
r_t = \frac{\bra{K^0 \kbar} H_d \ket{\bd}}{\bra{K^0 \kbar} H_s
  \ket{\bs} } = 1 + C_{\sss SU(3)} ~,~~
r_t^* = \frac{\bra{K^* \kstarbar} H_d \ket{\bd}}{\bra{K^*
\kstarbar} H_s \ket{\bs} } = 1 + C_{\sss SU(3)}^* ~,
\eeq
where $C_{\sss SU(3)}$ and $C_{\sss SU(3)}^*$ are both expected to be
$\sim 25\%$ (i.e.\ the typical size of $SU(3)$-breaking
effects). Thus,
\beq
\frac{r_t}{r_t^*} = 1 + (C_{\sss SU(3)} - C_{\sss SU(3)}^*) ~.
\label{corrections}
\eeq
Now, apart from $SU(3)$, there is no symmetry limit in which $(C_{\sss
SU(3)} - C_{\sss SU(3)}^*) \to 0$, so that one might guess that the
$SU(3)$ corrections to $r_t/r_t^*$ are also $\sim 25\%$. However, as
we argue below, we expect significant cancellations between $C_{\sss
SU(3)}$ and $C_{\sss SU(3)}^*$.

At the quark level, the $SU(3)$ breaking vanishes in the limit
$m_b\to\infty$, so that the hamiltonians describing the decays
$\bd \to K^0 \kbar$ and $\bs \to K^0 \kbar$ are equal to
$O(\Delta M_B
/ M_B) \simeq 2\%$.
Writing
\bea
r_t & = & \bra{K^0\kbar}H_d\ket{\bd}/\bra{K^0\kbar}H_s\ket{\bs} \nn\\
    & = & \bra{K^0\kbar}H_d\ket{\bd}/\bra{K^0\kbar}U^\dagger H_d U\ket{\bs} ~,
\eea
we see that there are two main sources of $SU(3)$-breaking
corrections: (i) ``final-state'' corrections, $U \ket{K^0\kbar} \ne
\ket{K^0\kbar}$, and (ii) ``initial-state'' corrections, $U\ket{\bs}
\ne \ket{\bd}$.

The key observation here is that the sources of $SU(3)$ breaking in
$r_t^*$ are very similar to those in $r_t$: $U \ket{K^* \kstarbar} \ne
\ket{K^* \kstarbar}$ and $U\ket{\bs} \ne \ket{\bd}$. It is therefore
not unreasonable to expect sizeable cancellations between $C_{\sss
SU(3)}$ and $C_{\sss SU(3)}^*$ in Eq.~(\ref{corrections}), and indeed
this is what is found in model calculations. Such calculations suggest
that both the final-state and initial-state corrections are at the
level of 1--2\%. Furthermore, this can be tested experimentally. For
the final-state corrections, one needs to measure the kaon light-cone
distribution at the scale of $m_b$, while information about the size
of initial-state corrections can be obtained from measurements of the
$D,D_s \to K, K^*$ form factors. For all the details concerning the
size of the theoretical uncertainty, as well as the experimental
tests, we refer the reader to Ref.~\cite{BKKbar}.

To summarize, we have presented a new method for obtaining the CP
phase $\alpha$ via measurements of $B^0_{d,s} \to K^{(*)} {\bar
K}^{(*)}$ decays. This method is particularly appropriate for hadron
colliders since some of the branching ratios are small [$O(10^{-6})$],
and since $\bs$ decays are involved. Furthermore, the final-state
particles can be detected via their decays to charged particles only;
no $\pi^0$ detection is needed. By comparing the value of $\alpha$
extracted from this method with that obtained in $B\to\pi\pi$ or
$B\to\rho\pi$ decays, one can detect the presence of new physics in
the $b\to d$ penguin amplitude. The method does require theoretical
input. However, model calculations suggest that the theoretical
uncertainty is at most 5\%, and might well be even smaller.
Furthermore, these estimates can be tested experimentally. The method
is therefore quite clean.

\bigskip
D.L. is thanks the organizers of FPCP2002 for a wonderful conference.
This work was financially supported by NSERC of Canada.

\end{document}